# Efficient quantum-chemical geometry optimization and the structure of large icosahedral fullerenes

Brett I. Dunlap

Code 6189, Theoretical Chemistry Section

US Naval Research Laboratory

Washington, DC 20375-5342

and

Rajendra Zope

Department of Chemistry,

George Washington University,

Washington DC, 20052

**Abstract**

Geometry optimization is efficient using generalized Gaunt coefficients, which significantly limit the amount of cross differentiation for multi-center integrals of high-angular-momentum solid-harmonic basis sets. The geometries of the most stable $C_{240}$, $C_{540}$, $C_{960}$, $C_{1500}$, and $C_{2160}$ icosahedral fullerenes are optimized using analytic density-functional theory (ADFT), which is parameterized to give the experimental geometry of $C_{60}$. The calculations are all electron, the orbital basis set includes *d* functions and the exchange-correlation-potential basis set includes *f* functions. The largest calculation on $C_{2160}$ employed about 39000 basis functions.

To our knowledge the geometry of $C_{240}$ has not been optimized using excellent basis sets, despite continuing interest in the problem [1]. Based on the $N^3$ scaling of the eigenvalue problem or the number of non-zero linear combination of atomic orbitals (LCAO) matrix elements involved, one would correctly estimate that accurate geometry optimization of $C_{2160}$ requires slightly less than a thousand times the resources required to optimize $C_{240}$. Such calculations are not totally impossible, however. A practical approach is analytic density-functional theory (ADFT) [2-4]. Its variational parameters are 10's of linear combination of atomic orbitals (LCAO) coefficients rather than 100's of plane-wave coefficients or numerical values at 1000's of points per atom per molecular orbital. A second step forward is to compute the contribution of the local density-of-states to forces not by recursion [5], and *xyz*-factorization [6] but via the generalized Gaunt coefficients [7]. Due to the present limitations [8] on the functional forms that can be treated in ADFT, however, we need to parameterize it in order to get the correct geometry of the giant fullerenes.

ADFT requires no numerical grid at all. Therefore matrix elements and thus the total energy can be computed to machine precision [9,10]. Furthermore, exact matrix elements mean that degeneracy can be removed from the problem. Both facts make accurate high-symmetry calculations much more practical than one might think. In ADFT the degeneracy of multidimensional irreducible representations (irreps) is entirely removed except when invoking Pitzer's theorem [11], which says that the sum of the magnitude squared of the basis functions for each irrep is invariant. That theorem means that self-consistent-field (SCF) integrals [12] and now forces need only be evaluated for symmetry-distinct bra-ket atomic pairs of LCAO basis functions. This efficiency enabled the largest *ab initio* SCF calculation on icosahedral $C_{240}$ [13], which is much more stable than the icosahedral $C_{60}$ cluster that can be made dominant among the fullerenes [14,15], for reasons that are still not completely understood.

Efficient computation requires switching from the traditional Cartesian-Gaussian basis [5] to the solid-harmonic-Gaussian basis, which minimally contains all essential chemistry; the latter are eigenstates of angular momentum as are the atomic orbitals that collectively they approximate. The matrix elements corresponding to higher angular momentum are computed by differentiating the *s*-type matrix elements with respect to the corresponding atomic center [9]. Angular momentum increases one unit with each appropriate differentiation. The solid-harmonics of nabla [16] create angular momentum and are generated recursively by a two-term expression,

$$(L-M)[\nabla]_M^L = \frac{\partial}{\partial z}[\nabla]_M^{L-1} + (-\frac{\partial}{\partial x} + i\frac{\partial}{\partial y})[\nabla]_{M+1}^{L-1}, \tag{1}$$

and its complex conjugate, using $[\nabla]_M^{L*} = (-)^M [\nabla]_{-M}^L$, and $[\nabla]_0^0 = 1$. Differentiation of solid-harmonics maximally lowers their angular momentum [17], and thus they are closed under differentiation. The solid-harmonic addition theorem extends the product rule of differentiation to these differential operators [16]. Thus no amount of solid-harmonic differentiation creates anything other than solid harmonics which are conveniently described as angular momentum about the various molecular centers. The generalized Gaunt coefficients arise when an *s*-type function is operated upon by multiple solid-harmonic differential operators. Angular momentum is lost when a differential operator acts upon a solid harmonic created by another operator. The generalized Gaunt coefficients restricts that loss to zero total angular momentum. Their use becomes more efficient as the angular momentum on each basis function is increased.

In our normalization, Eq.1, the addition theorem is factorless [18], and thus independent of initial angular momentum, which means that the entire reduced density matrix for each symmetry-distinct pair of centers can be summed for each set of angular momentum lost by those two centers. In our code this sum is repeated for each third center of the Kohn-Sham potential [19]. This approach tremendously speeds up analytic derivatives of three-center integrals via the 4-*j* generalized Gaunt coefficient [20]. This sum over geometric factors is still much work; it becomes 37% of the entire time it takes to optimize our largest fullerene. Under its direct-SCF option, our code becomes much slower and this percentage drops accordingly. The code is quite scalable as well as efficient when each processor reads its share of the three-center integrals from its own disk.

For the single-element case of the fullerenes, our code is identical to the SCF code of Werpetinski and Cook [21]. This method requires a Gaussian basis, which we choose to be 6-311-G* [22], with highest angular momentum *d*-type, for analytically and variationally fitting the orbitals. We also need auxiliary bases for analytically and variationally fitting the charge density, its cube root, and its cube root squared. For the *s*-type components of all three fitting bases we uncontract and appropriately scale the *s*-type orbital exponents [2]. All higher angular momentum fitting functions are the uncontracted and unscaled density-fitting exponents from an optimization, the highest angular-momentum component of which is *f*-type [23]. There are 18 orbital basis functions per atom and at most 42 fitting functions (of three types) per symmetry inequivalent atom.

The best standard method [24] can only converge the SCF calculations to the point where the maximum iteration-to-iteration occupied-virtual overlap matrix element is $10^{-10}$. The off-diagonal 1s-1s iteration-to-iteration overlaps in each symmetry block are up to three orders of magnitude larger. This instability, perhaps associated with our single-$\zeta$ treatment of the 1s cores, limits our SCF convergence, which we measure by largest change in a density-fitting coefficient. We had to increase the allowed change to $10^{-6}$ in the coefficient of a charge-density basis function with unit Coulomb self-interaction. With this limitation we are able to optimize the structure of the special fullerenes to the point where the root-mean-square gradient is less than $10^{-4}$ Hartree/Bohr.

In the infinite limit a fullerene is a graphene sheet closed by 12 pentagons, which negligibly affects many properties including the binding energy and median nearest-neighbor bond distance. Total *ab-initio* energies, which can quite accurately be evaluated at empirical geometries due to the variational principle, as a function of fullerene size showed that band-structure calculations significantly underestimated the DFT binding energy of graphene and graphite [25] and lead to better band-structure calculations on graphite [26]. We can now just as rapidly compute fullerene geometries within ADFT for X$\alpha$ exchange functionals. The early appeal of the X$\alpha$ method was that it is the unique quantum chemical method which allows molecules to dissociate correctly into atoms. This property leads to excellent total molecular energies [4]. Perhaps optimizing Slater's exchange parameter, $\alpha$ [27], can be used to extrapolate other molecular properties. We used Perl scripts to determine the $\alpha$ value, 0.684667, that gives the experimental bond distances of $C_{60}$. The geometries using this $\alpha$ are given in Table I. The coordinate axes are two-fold symmetry axes, thus an atom with a zero coordinate is one of sixty symmetry-equivalent atoms otherwise each tabulated atom generates 120 others. The median nearest-neighbor bond distance for each fullerene is given in Table II. The median bond distance might be going to that of graphite, which is 1.422 [26]. Certainly the mean bond distance for large fullerenes and graphene is less than 0.1% too long with this value of $\alpha$ in ADFT. The standard deviation of the radius of each atom from the fullerene's center is also given in Table II and compared with that of a tight-binding calculation [28]. Our structures are slightly more facetted. As these are most likely the most accurate fullerene geometries available, their complete geometries can be constructed from the coordinates given in Table 1.

With reliable structures, one can consider one-shot energy calculations with methods of higher accuracy, but for which geometry optimization is impractical at the moment. Another alternative is to develop empirical methods for energy evaluation. More in line with the second approach we have determined the value $\alpha$ value, 0.64190, that gives the experimental atomization energy of $C_{60}$, which we take as 7.14 eV, which corresponds to an enthalpy of formation of 0.43 eV/atom [29] relative to graphite's 7.37 eV/atom [30], ignoring zero-point energy differences. To compute the total energy of our reference carbon atom we use $C_{2v}$ symmetry for which spin-polarized ADFT gives integral occupation numbers. With this technology we estimate the atomization energy per carbon atom of these fullerenes in the final column of Table II. This result is disappointing because already by $C_{240}$ the atomization energy per atom is greater than that of graphite. Clearly better ADFT energy functionals are needed, but those that we now have allow large basis-set calculations on some very large systems and can apparently give reliable geometries. This work shows that standard convergence and optimization methods are sufficient for molecules as large as $C_{2160}$.

**Acknowledgment**

We thank Prof. Peter Pulay for discussions about convergence. The Office of Naval Research, directly and through the Naval Research Laboratory, and the DoD's High Performance Computing Modernization Program, through the Common High Performance Computing Software Support Initiative, Project MBD-5, supported this research.

Table I. Coordinates in Angstroms of the symmetry-inequivalent carbon atoms in icosahedral fullerenes in a coordinate system in which the coordinate axes are two-fold symmetric. Thus atoms in a coordinate axis plane are equivalent to 60 rather than 120 other atoms.

| $C_{60}$ | | |
|---|---|---|
| 3.4785 | 0.6991 | 0.0000 |
| $C_{240}$ | | |
| 6.7860 | 0.7115 | 1.2648 |
| 6.9303 | 1.3919 | 0.0000 |
| 6.7985 | 2.7763 | 0.0000 |
| $C_{540}$ | | |
| 10.3325 | 3.5198 | 0.0000 |
| 10.2803 | 0.7073 | 0.0000 |
| 10.2436 | 4.9086 | 0.0000 |
| 10.1695 | 2.8338 | 1.2476 |
| 10.1779 | 1.4142 | 1.2463 |
| 5.1271 | 6.9966 | 5.3981 |
| $C_{960}$ | | |
| 3.5432 | 5.8199 | 11.8037 |
| 4.9703 | 3.5496 | 12.7208 |
| 2.8466 | 4.7639 | 12.4593 |
| 5.6793 | 2.4027 | 13.1633 |
| 1.4252 | 4.8034 | 12.5708 |
| 3.5573 | 3.6032 | 12.8776 |
| 1.4201 | 0.0000 | 13.6515 |
| 2.8364 | 0.0000 | 13.6734 |
| 7.0406 | 0.0000 | 13.6930 |
| 5.6502 | 0.0000 | 13.7765 |
| | $C_{1500}$ | |
| 3.5512 | 7.0518 | 15.0185 |
| 7.8104 | 2.4025 | 16.6105 |
| 4.9511 | 6.9501 | 14.7778 |
| 5.6653 | 5.8047 | 15.2231 |
| 4.9761 | 4.7588 | 15.8998 |
| 5.6878 | 3.6005 | 16.3197 |
| 7.1003 | 3.5465 | 16.1622 |
| 1.4247 | 5.9997 | 15.7080 |
| 2.8479 | 5.9672 | 15.6202 |
| 3.5577 | 4.8042 | 16.0266 |
| 4.9666 | 0.0000 | 17.1112 |
| 3.5497 | 0.0000 | 17.0721 |
| 7.7814 | 0.0000 | 17.2247 |
| 9.1725 | 0.0000 | 17.1423 |
| 0.7093 | 0.0000 | 17.0074 |
| | $C_{2160}$ | |
| 18.0082 | 4.9622 | 8.1894 |
| 18.4625 | 5.6791 | 7.0479 |
| 7.1054 | 4.7521 | 19.3344 |
| 7.8192 | 3.6012 | 19.7701 |
| 9.2300 | 3.5423 | 19.6017 |
| 9.9418 | 2.4034 | 20.0606 |
| 7.0731 | 6.9341 | 18.1973 |
| 7.7955 | 5.8055 | 18.6713 |
| 17.2983 | 5.6441 | 9.2173 |
| 18.7696 | 2.8471 | 7.1695 |
| 19.1851 | 3.5591 | 6.0083 |
| 19.0625 | 4.9781 | 5.9624 |
| 5.6898 | 4.8044 | 19.4785 |
| 18.2145 | 3.5547 | 8.2754 |
| 18.8481 | 1.4255 | 7.1994 |
| 1.4202 | 0.0000 | 20.3958 |
| 2.8390 | 0.0000 | 20.4273 |
| 7.0968 | 0.0000 | 20.5677 |
| 5.6803 | 0.0000 | 20.4983 |
| 9.9124 | 0.0000 | 20.6708 |

| 11.3039 | 0.0000 | 20.5910 |

Table II.  The median nearest-neighbor bond distance, average radius, radial standard deviation, all in Angstroms, for the fullerenes of this work computed using $\alpha = 0.684667$. The average radii and radial standard deviations are slight larger than a published tight-binding (TB) calculation [28].  The right-hand column gives the atomization energy, in electron volts, that we compute using $\alpha = 0.64190$.

| Fullerene | Median Bond Distance | Average Radius | TB Average Radius [28] | Radial Standard Deviation | TB Radial Standard deviation [28] | Atomization Energy/atom ($\alpha = 0.64190$) |
|---|---|---|---|---|---|---|
| $C_{60}$   | 1.4244 | 3.5481  |       | 0.000 |      | -7.140 |
| $C_{240}$  | 1.4306 | 7.0728  | 7.06  | 0.165 | 0.15 | -7.373 |
| $C_{540}$  | 1.4264 | 10.5528 | 10.53 | 0.360 | 0.35 | -7.431 |
| $C_{960}$  | 1.4249 | 14.0342 | 14.02 | 0.526 | 0.52 | -7.459 |
| $C_{1500}$ | 1.4244 | 17.5225 |       | 0.677 |      | -7.474 |
| $C_{2160}$ | 1.4241 | 21.0137 | 20.95 | 0.822 | 0.82 | -7.484 |

**References:**

[1] G. E. Scuseria, Science **271**, 942 (1996).

[2] B. I. Dunlap, J. Phys. Chem. A **107**, 10082 (2003).

[3] R.R. Zope and B.I. Dunlap, Chem. Phys. Lett. **399**, 417 (2004)

[4]R.R. Zope and B.I. Dunlap, Phys. Rev. B **71**, 193104 (2005)

[5]  Y. Shao, C. A. White, and M. Head-Gordon, J. Chem. Phys. **114**, 6572 (2001).

[6]  D. Sundholm, J. Chem. Phys. **122**, 194107 (2005).

[7] B. I. Dunlap, Phys. Rev. A **66**, 032502 (2002)

[8] R. R. Zope and B. I. Dunlap, J. Chem. Phys., **124**, 044107 (2006).

[9] S. F. Boys, Proc. Roy. Soc. **200**, 542 (1950).

[10] J. A. Pople, Angw. Chem. Int. Ed. **38**, 1894 (1999).

[11] R. M. Pitzer, J. Chem. Phys. **58**, 3111 (1973).

[12] B. I. Dunlap, Adv. Chem. Phys. **69**, 287 (1987).

[13] B.I. Dunlap, D.W. Brenner, J.W. Mintmire, R.C. Mowrey, and C.T. White, J. Phys. Chem. **95**, 8737-8741 (1991)

[14] H. W. Kroto, J. R. Heath, S. C. O'Brien, R. F. Curl, and R. E. Smalley, Nature **318**, 162 (1985).

[15] W. Krätschmer, L. D. Lamb, K. Fostiropoulos, and D. R. Huffman, Nature, **347**, 354 (1990).

[16] B. I. Dunlap, Phys. Rev. A **42**, 1127 (1990).

[17] E. J. Weniger, Collect. Czech. Chem. Comm. **70**, 1225 (2005).

[18] E. O. Steinborn and K. Ruedenberg, Adv. Quantum Chem. **7**, 1 (1973).

[19] B. I. Dunlap, Comp. Phys. Commun. **165**, 18 (2005).

[20] B. I. Dunlap, Int. J. Quantum. Chem. **81**, 373 (2001).

[21] K. S. Werpetinski and M. Cook, Phys. Rev. A **52**, R3397 (1995); J. Chem. Phys. **106**, 7124 (1997).

[22] R. Krishnan, J. S. Binkley, R. Seeger, and J. A. Pople, J. Chem. Phys. **72**, 650 (1980).

[23] K. Eichkorn, O. Treutler, H. Öhm, M. Häser, R. Ahlrichs, Chem. Phys. Lett. **240**, 283 (1995).

[24] P. Pulay, J. Comp. Chem. **3**, 556 (1982).

[25] B.I. Dunlap and J.C. Boettger, J. Phys. B: At. Mol. Opt. Phys. **29**, 4907 (1996).

[26] J. C. Boettger, Phys. Rev. B **55**, 11202 (1997).

[27] J. C Slater, *Quantum Theory of Molecules and Solids,* Vol. IV (McGraw-Hill, New York, 1974).

[28] S. Itoh, P. Ordejón, D. A. Drabold, and R. M. Martin, Phys. Rev. B **53**, 2132 (1996).

[29] H. P. Diogo, M. E. Mineas de Piedade, T. J. S. Dennis, J.P. Hare, H. W. Kroto, R. Taylor, and D. R. M. Watson, J. Chem. Soc.-Faraday Trans. **89**, 3541 (1993).

[30] C. Kittel, *Introduction to Solid State Physics* 5th Edn. (New York, Wiley, 1976) p. 74.